\documentclass{notices}
\usepackage{amsfonts,amssymb,amsmath,amscd,enumerate}
\usepackage{float}
\usepackage{makeidx}
\usepackage{graphicx}
\usepackage{geometry}
\usepackage{mathtools}

\title{Fox's $H$-Functions:
 A Gentle Introduction Through Astrophysical Thermonuclear Functions
}

\author{Hans J. Haubold
	\affil{Office for Outer Space Affairs, Vienna International Centre, Austria hans.haubold@gmail.com.
	}\and
Dilip Kumar
	\affil{Department of Mathematics, University of Kerala, Kariavattom, India ashikak@keralauniversity.ac.in
	}\and 	Ashik A. Kabeer
	\affil{Department of Mathematics, University of Kerala, Kariavattom, India dilipkumar@keralauniversity.ac.in}}

\begin{document}

\maketitle

\section{Abstract}
Needed for cosmological and stellar nucleosynthesis, we are studying the closed form analytic evaluation of thermonuclear reaction rates. In this context, we undertake a comprehensive analysis of three distinct velocity distributions, namely the Maxwell-Boltzmann distribution, the pathway distribution, and the Mittag-Leffler distribution. We emphasize the utilization of Meijer $G$-function and Fox $H$-function which are special functions of mathematical physics.

\section{$H$- function}
The invention of $G$ and $H$-functions has created a revolution in the special function theory, as these are two functions that are general in nature, covering many genesis of special functions, and have wide applicability. C.S. Meijer in 1936 \cite{Meijer1936} introduced the $G$-function as a generalisation of hypergeometric function in terms of the Mellin-Barnes contour integral. Following the definition of Meijer's $G$-function, in 1961, Charles Fox \cite{fox1961} defined a new function involving the Mellin-Barnes integrals covering Meijer's $G$-function, which he called $H(x)$, later known as the $H$-function. The $H$-function turns out to be a generalisation of many known special functions at that time, namely hypergeometric functions, wright functions,  Mittag-Leffler functions, Bessel functions, $G$-functions, etc.
\par 
The $H$-function is defined for integers $m,n,q,r$ with $0\leq m\leq q$, $0\leq n\leq r$, $c_i,~d_j\in\mathbb{C}$ and for $C_i$, $D_j$ $\in\mathbb{R_{+}}=(0,\infty)$ by
\begin{align}\label{H}
	H_{r,q}^{m,n}&\begin{pmatrix}
		z\bigg\rvert
		\begin{matrix*}
			(c_1,C_1),\dots,(c_r,C_r)\\
			(d_1,D_1),\dots,(d_q,D_q)
		\end{matrix*}
	\end{pmatrix}\nonumber\\&:=\displaystyle\frac{1}{2\pi i}\int_{\mathcal{L}}\Phi(s)z^{-s}{\rm d}s
\end{align}
\begin{align}
	\Phi(s)&=\displaystyle\frac{\left\{\prod \limits_{j=1}^{m}\Gamma(d_{j}+D_js)\right \}\left\{\prod \limits_{j=1}^{n}\Gamma(1-c_{j}-C_js)\right \}}{\left\{\prod \limits_{j=m+1}^{q}\Gamma(1-d_{j}-D_js)\right \}\left\{\prod \limits_{j=n+1}^{r}\Gamma(c_{j}+C_js)\right \}}\nonumber
\end{align}
where
$$z^{-s}=\exp[-s\{\ln |z| + \arg z\}]; z\neq  0; i = \sqrt{-1};$$
where $\arg(z)$ is not necessarily the principal value. Here $\mathcal{L}$ is an appropriate contour separating the poles 
$$\zeta_jl=-\left(\frac{d_j+l}{D_j}\right),j=1,2,\ldots,m;l=0,1,2,\ldots$$
of $\Gamma(d_j+D_js),$ $j=1,\dots, m$ from those poles 
$$\eta_ik=-\left(\frac{1-c_i+k}{C_i}\right),i=1,2,\ldots,n;k=0,1,2,\ldots$$
of $\Gamma(1-c_i-C_is)$ $i=1,\dots,n$. Any empty product is considered as unity and it is assumed that no poles of $\Gamma(d_j+D_js)$ for $j=1,2,...,m$ coincides with any poles of $\Gamma(1-c_j-C_j)$ for $j=1,2,...,n$, where $\Gamma(\cdot)$ is the Euler gamma function. The details on the different types of contours for $\mathcal{L}$ and the convergence conditions for the existance is available in numerous books, see for example Mathai et al. \cite{HH}, Kilbas and Saigo \cite{kil}, Prudnikov et al. \cite{pr}, etc. 
It is evident that  the Mellin transform of  $H_{r,q}^{m,n}\begin{pmatrix}
	z\bigg\rvert
	\begin{matrix*}
		(c_1,C_1),\dots,(c_r,C_r)\\
		(d_1,D_1),\dots,(d_q,D_q)
	\end{matrix*}
\end{pmatrix}$ is $\Phi(s)$.
Let \begin{align}
	\Delta&=\sum\limits_{j=1}^{q}D_j-\sum\limits_{i=1}^{r}C_i\nonumber\\
	a^{*}&=\sum\limits_{i=1}^{n}C_i-\sum\limits_{i=n+1}^{r}C_i+\sum\limits_{j=1}^{m}D_j-\sum\limits_{j=n+1}^{q}D_j\nonumber\\
	\mu^*&=\sum\limits_{j=1}^{q}d_j-\sum\limits_{i=1}^{r}c_i+\frac{r-q}{2}\nonumber\\
	\delta^*&=\prod\limits_{i=1}^{r}C_i^{-C_i}\prod\limits_{j=1}^{q}D_j^{D_j},\nonumber
\end{align}then the existence conditions for Fox's $H$-function is given by the following cases \cite{h}:

\begin{align}
	1.~\mathcal{L}&=\mathcal{L}_{-\infty}, ~\Delta>0,~z\neq 0,\nonumber\\
	2.~\mathcal{L}&=\mathcal{L}_{-\infty}, ~\Delta=0, ~0<|z|<\delta^*,\nonumber\\ 	3.~\mathcal{L}&=\mathcal{L}_{-\infty}, ~\Delta=0,~|z|=\delta^*,\Re(\mu^*)<-1\nonumber,
\end{align}where $\mathcal{L}_{-\infty}$ is a left loop starting at $-\infty+i\lambda_1$ and terminating at $-\infty+i\lambda_2$, $-\infty<\lambda_1<\lambda_2<\infty.$
\begin{align}
	4.~\mathcal{L}&=\mathcal{L}_{+\infty}, ~\Delta<0,~z\neq 0,\nonumber\\
	5.~\mathcal{L}&=\mathcal{L}_{+\infty}, ~\Delta=0, ~|z|>\delta^*,\nonumber\\ 	6.~\mathcal{L}&=\mathcal{L}_{+\infty}, ~\Delta=0,~|z|=\delta^*,\Re(\mu^*)<-1\nonumber,
\end{align}where $\mathcal{L}_{+\infty}$ is a right loop starting at $+\infty+i\lambda_1$ and terminating at $+\infty+i\lambda_2$, $-\infty<\lambda_1<\lambda_2<\infty.$
\begin{align}
	7.~\mathcal{L}&=\mathcal{L}_{i p\infty}, ~a^*>0,~|\arg z|<\frac{a^*\pi}{2}\nonumber\\
	8.~\mathcal{L}&=\mathcal{L}_{i p\infty}, ~a^*=0,\nonumber\\&\Delta p+\Re(\mu^*)<-1,~\arg z=0,z\neq0 \nonumber,
\end{align}where $\mathcal{L}_{ip\infty}$ is a contour starting at $p-i\infty$ and terminating at $p+i \infty$, $p\in \mathbb{R}.$

\begin{figure}[!htb]
	\begin{minipage}{0.5\textwidth}
		\centering
		\includegraphics[width=1\linewidth]{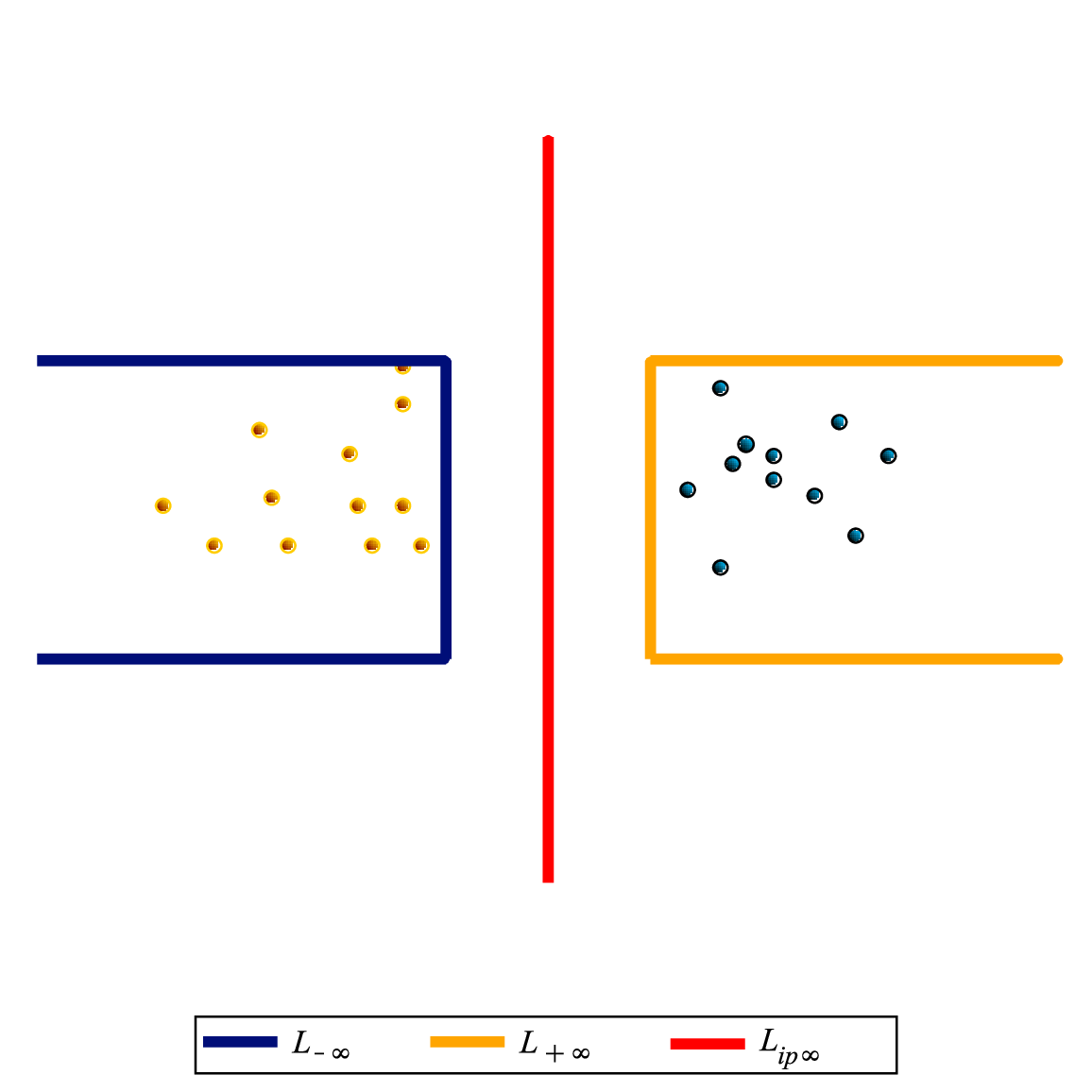}
		\caption{Types of contours for $H$-function}\label{Fig:Data0}
	\end{minipage}\hfill
\end{figure}\par For instance, set $C_i$, $D_j=1$ in above definition of $H$-function, we obtain the Meijer $G$-function, defined as\begin{align}
G_{r,q}^{m,n}&\begin{pmatrix}
	z\bigg\rvert
	\begin{matrix*}
		c_1,\dots,c_r\\
		d_1,\dots,d_q
	\end{matrix*}
\end{pmatrix}\nonumber\\&:=\displaystyle\frac{1}{2\pi i}\int_{\mathcal{L}}\phi(s)z^{-s}{\rm d}s\nonumber
\end{align}
\begin{align}
\phi(s)&=\displaystyle\frac{\left\{\prod \limits_{j=1}^{m}\Gamma(d_{j}+s)\right \}\left\{\prod \limits_{j=1}^{n}\Gamma(1-c_{j}-s)\right \}}{\left\{\prod \limits_{j=m+1}^{q}\Gamma(1-d_{j}-s)\right \}\left\{\prod \limits_{j=n+1}^{r}\Gamma(c_{j}+s)\right \}}\nonumber,
\end{align}where $\mathcal{L}$ is the suitable contour. The existence conditions for $G$-function can be found from the above existence condition for $H$-function.

{\section{Thermonuclear reaction}}
\par The sun and other stars are governed by explosive thermonuclear fusion processes.  These reactions involve the collision of atomic nuclei at incredibly high temperatures, typically millions of degree Celsius, found in a star's core. At such extreme conditions, the velocity distribution of the interacting particles is governed by the principles of plasma physics. For those charged particles, the Maxwell-Boltzmann velocity distribution describes how particles within the plasma move and collide. Understanding this distribution is crucial, as it dictates the likelihood of particles having sufficient kinetic energy to overcome the electrostatic repulsion and engage in nuclear fusion. Essentially, the velocity distribution of particles drives the reaction rates and probabilities of thermonuclear reactions, determining the star's energy output and, ultimately, its lifespan.
\par In a non-degenerate setting, the particles of type $i$ and $j$ have the reaction rate $r_{ij}$ given by \cite{hauboldkumar2008}, 
\begin{equation}\label{01}
	r_{ij}=n_in_j\langle\sigma_{ij}v\rangle 
\end{equation} 
\noindent	where $n_i$ and $n_j$ are the particle number densities of particles of type $i$ and $j$ respectively and $\langle\sigma_{ij}v\rangle$ is an appropriate average of the product of the energy-dependent reaction cross-section and relative velocity of the interacting particles:
\begin{align*}
	\langle\sigma_{ij}v\rangle=\int_{0}^{\infty}vf(v)\sigma(v){\rm d}^3v,
\end{align*} where $\sigma(v)$ is the reaction cross section, $f(v)$ is the distribution function of the relative velocity of the particles $i,~j$ and ${\rm d}^3v=4\pi v^2 {\rm d}v.$
{\subsection{Maxwell-Boltzmann Case}}

\par The Maxwell-Boltzmann distribution is a fundamental concept that describes the distribution of particle velocities within a high-temperature environment, such as the core of a star. In this distribution, particles like protons and helium nuclei exhibit a wide range of velocities due to their thermal motion. Some particles possess energies and velocities high enough to overcome the Coloumb barrier and successfully collide, initiating nuclear fusion. The Maxwell-Boltzmann distribution helps us understand the probability of different particles having the necessary kinetic energy for fusion events, which is vital for comprehending the rates and efficiencies of thermonuclear reactions that power the stars and drive the energy balance within celestial objects. For a non-relativistic,
non-degenerate plasma of nuclei in thermodynamic equilibrium, the particles in the plasma
possess a classical Maxwell–Boltzmann velocity distribution given by \cite{anal}
\begin{align}\label{v1}
	f_{MBD}(v){\rm d}v=\left(\frac{\mu}{2 \pi KT} \right)  ^{\frac{3}{2}}e^{-\frac{\mu v^2}{2KT} } 4 \pi v^2{\rm d}v, 
\end{align}where $\mu$ is the reduced mass of reacting particles, $K$ is the Boltzmann's constant, $T$ is the temperature.
In terms of relative kinetic energy, the Maxwell-Boltzmann energy distribution of the interacting particles is given by
\begin{align}
	f_{MBD}(E){\rm d}E=2\pi \displaystyle\left(  \frac{1}{\pi KT}\right) ^{\frac{3}{2}}e^{-\frac{E}{KT} } \sqrt{E}{\rm d}E,\label{456}
\end{align}where $E=\frac{\mu v^2}{2},$ Now, the reaction rate  in the Maxwell-Boltzmann case can be obtained as
\begin{align}
	&~~~{r}_{ij}=n_in_j\left(\frac{8 }{\mu\pi} \right)^{\frac{1}{2}}\left(\frac{1}{ KT} \right)^{\frac{3}{2}}\sum\limits_{\nu=0}^{2}\frac{S^{(\nu)}(0)}{\nu!}\nonumber\\&\times\int_{0}^{\infty}E^{\nu}\exp {\left[-\frac{E}{KT}-2\pi\left(\frac{\mu}{2} \right)^{\frac{1}{2}} \frac{z_iz_j{\rm e}^2}{\hbar E^{\frac{1}{2}}}  \right]}{\rm d}E\nonumber
\end{align} which in terms of Meijer's $G$- function is
\begin{align}
	&~~~{r}_{ij}=\frac{n_in_j}{\pi}\left(\frac{8 }{\mu} \right)^{\frac{1}{2}}\sum\limits_{\nu=0}^{2}\frac{S^{(\nu)}(0)}{\nu!}\left(\frac{1}{ KT} \right)^{-\nu+\frac{1}{2}}\nonumber\\&\times G_{0,3}^{3,0}\begin{pmatrix}
		\frac{x^2}{4}&\bigg\lvert
		\begin{matrix*}
			-- \\
		0,\frac{1}{2},\nu+1
		\end{matrix*}
	\end{pmatrix},\nonumber
\end{align} where $x=2\pi\left(\frac{\mu}{2} \right)^{\frac{1}{2}} \frac{z_iz_j{\rm e}^2}{\hbar},$ ${\rm e}$ is the quantum of electric charge, $\hbar$ is the Plank’s quantum of action.
\par One might imagine an alternate velocity distribution in place of the Maxwell-Boltzmann distribution for nuclear fusion processes occurring in star interiors if there is a deviation from hydrostatic equilibrium. The following sections discusses the alternative distributions which also covers the classical Maxwell-Boltzmann distribution.
{\subsection{Pathway Case}}

\par Using the principle of pathway model in real scalar case \cite{2007}, Haubold and Kumar \cite{hauboldkumar2008} extended the non-resonant reaction rate integrals, covering the non-extensive statistical mechanics of Tsallis' \cite{tsallis}. Also represented the extended reaction rates in the form of Fox's $H$-function. The Pathway energy distribution considered by Haubold and Kumar \cite{hauboldkumar2008} is given by 
\begin{align}\label{457}
	f_{PD}(E){\rm d}E&=\displaystyle \frac{2\pi (\alpha-1)^{\frac{3}{2}}}{(\pi KT)^{\frac{3}{2}}}\frac{\Gamma\left(\frac{1}{\alpha-1} \right) }{\Gamma\left(\frac{1}{\alpha-1}-\frac{3}{2} \right)}\nonumber\\&\times\sqrt{E}\left[1+(\alpha-1)\frac{E}{KT}\right]^{-\frac{1}{\alpha-1}}{\rm d}E,
\end{align}where $\alpha>1, ~\frac{1}{\alpha-1}>\frac{3}{2}$. Using (\ref{01}), the reaction rate in pathway case is given by\begin{align}
~~~&{r}_{ij}=n_in_j\left(\frac{8 }{\mu \pi} \right)^{\frac{1}{2}} \frac{ (\frac{\alpha-1}{{KT}})^{\frac{3}{2}}\Gamma\left(\frac{1}{\alpha-1} \right) }{\Gamma\left(\frac{1}{\alpha-1}-\frac{3}{2} \right)}\nonumber\\&\times\sum\limits_{\nu=0}^{2}\frac{S^{(\nu)}(0)}{\nu!}\int_{0}^{\infty}E^{\nu}\left[ 1+(\alpha-1)\frac{E}{KT}\right] ^{-\frac{1}{\alpha-1}}\nonumber\\&\times\exp {\left[-2\pi\left(\frac{\mu}{2} \right)^{\frac{1}{2}} \frac{z_iz_j{\rm e}^2}{\hbar E^{\frac{1}{2}}}  \right]}{\rm d}E,
\end{align}which in closed form gives 
\begin{align}
	~~~{r}_{ij}&=n_in_j\left(\frac{8 }{\mu \pi} \right)^{\frac{1}{2}} \frac{ (\alpha-1)^{\frac{3}{2}} }{\Gamma\left(\frac{1}{\alpha-1}-\frac{3}{2} \right)}\nonumber\\&\times\sum\limits_{\nu=0}^{2}\left(\frac{1}{KT} \right)^{-\nu+\frac{1}{2}}\frac{S^{(\nu)}(0)}{\nu!}\frac{1}{(\alpha-1)^{\nu+1}}\nonumber\\&\times G_{1,3}^{3,1}\begin{pmatrix}
		\frac{x^2(\alpha-1)}{4}&\bigg\lvert
		\begin{matrix*}
			2-\frac{1}{\alpha-1}+\nu \\
			0,\frac{1}{2},\nu+1
		\end{matrix*}
	\end{pmatrix}.
	\end{align}The corresponding velocity distribution in the pathway extended case is given by
	
\begin{align}\label{v2}
	f_{PD}(v){\rm d}v&=\left(\frac{(\alpha-1)\mu}{2 \pi KT} \right)^{\frac{3}{2}} \frac{\Gamma\left(\frac{1}{\alpha-1} \right) }{\Gamma\left(\frac{1}{\alpha-1}-\frac{3}{2} \right)}\nonumber\\&\times \left[1+(\alpha-1)\frac{\mu v^2}{2KT}\right]^{-\frac{1}{\alpha-1}}4\pi v^2{\rm d}v.
\end{align}As $\alpha$ approaches to $1$, one can retrieve the Maxwell-Boltzmann case.
{\subsection{Mittag-Leffler Case}}

	\par Mittag-Leffler function, a special function of mathematical physics, can describe many complex behaviour and phenomena as it is a generalization of the exponential function. Using the idea of one-parameter standard Mittag-Leffler Gaussian distribution model introduced by Agahi and Alipour \cite[(2.2)]{Mi}, the {{M}}ittag-{{L}}effler velocity distribution can be obtained as
\begin{align}\label{v3}
	f_{ML}(v){\rm d}v=\frac{\Gamma\left(1-\frac{\beta}{2} \right) }{\beta \sqrt{\pi}}\left(\frac{\mu}{2\pi KT} \right)^{\frac{3}{2}}\nonumber \\
	\times~\mathbb{E}_{\beta,\beta}\left(-\frac{\mu v^2}{2KT} \right)4\pi v^2 {\rm d}v, 
\end{align}where $0<\beta\leq1$ and $\mathbb{E}_{\beta,\beta}(\cdot)$ is the Mittag-Leffler function,
\begin{align*}
	\mathbb{E}_{\alpha,\beta}(z)=\sum\limits_{k=0}^{\infty}\frac{z^k}{\Gamma(\alpha k
		+\beta)},~~z,~\beta\in \mathbb{C},~\Re(\alpha)>0.
\end{align*} In terms of the relative kinetic energy, the Mittag-Leffler energy distribution can be defined as,
\begin{align}\label{en}
	f_{ML}(E){\rm d}E=&\displaystyle \frac{2\sqrt{\pi }}{\beta}\Gamma\left(1-\frac{\beta}{2} \right)\left(\frac{1}{\pi KT} \right)^{\frac{3}{2}}\nonumber\\&\times\mathbb{E}_{\beta,\beta}\left(-\frac{E}{KT}\right) \sqrt{E} {\rm d}E,
\end{align}where $0<\beta\leq1.$ Now, using (\ref{01}), we obtain the reaction rate probability integral in the modified form as
\begin{align}
	&~~~{r}_{ij}=n_in_j\left(\frac{8\pi }{\mu} \right)^{\frac{1}{2}} \displaystyle \frac{\Gamma\left(1-\frac{\beta}{2} \right)}{\beta}\left(\frac{1}{\pi KT} \right)^{\frac{3}{2}}\nonumber\\&\times\sum\limits_{\nu=0}^{2}\frac{S^{(\nu)}(0)}{\nu!}\int_{0}^{\infty}E^{\nu}\mathbb{E}_{\beta,\beta}\left( -\frac{E}{KT}\right) \nonumber\\&\times\exp\left( {-2\pi\left(\frac{\mu}{2} \right)^{\frac{1}{2}} \frac{z_iz_j{\rm e}^2}{\hbar E^{\frac{1}{2}}} }\right) {\rm d}E.
\end{align} By putting $y=\frac{E}{KT}$ and $x=2\pi\left(\frac{\mu}{2} \right)^{\frac{1}{2}} \frac{z_iz_j{\rm e}^2}{\hbar}, $ we get the simplified form of the above integral as
\begin{align}
	{r}_{ij}&=n_in_j\left(\frac{8 }{\mu} \right)^{\frac{1}{2}} \displaystyle \frac{\Gamma\left(1-\frac{\beta}{2} \right)}{\beta \pi }\sum\limits_{\nu=0}^{2}\left(\frac{1}{ KT} \right)^{-\nu+\frac{1}{2}}\nonumber\\&\times\frac{S^{(\nu)}(0)}{\nu!}\int_{0}^{\infty}y^{\nu}\mathbb{E}_{\beta,\beta}\left(-y\right)e^{-xy^{-\frac{1}2}}{\rm d}y,\nonumber\end{align} which in terms of $H$-function is,\begin{align}
	&{r}_{ij}=n_in_j\left(\frac{8 }{\mu} \right)^{\frac{1}{2}} \displaystyle \frac{\Gamma\left(1-\frac{\beta}{2} \right)}{\beta \pi }\nonumber\\&\times\sum\limits_{\nu=0}^{2}\left(\frac{1}{ KT} \right)^{-\nu+\frac{1}{2}}\frac{S^{(\nu)}(0)}{\nu!z^{\nu+1}}\nonumber\\&\times H_{1,3}^{2,1}\begin{pmatrix}
		x&\bigg\lvert
		\begin{matrix*}
			\left( \nu+1,\frac{1}{2}\right)\\
			(0,1),~\left( \nu+1,\frac{1}{2}\right) ,~(\beta\nu+\beta,\frac{\beta}{2})
		\end{matrix*}
	\end{pmatrix}.\nonumber
\end{align}The detailed evaluation procedure for the above integral is provided in the Appendix. \par If there appear a cut-off in the high energy tail of the Mittag-Leffler modified distribution function, the reaction rate can be interpreted as 
\begin{align}
	{r}_{ij}&=n_in_j\left(\frac{8 }{\mu} \right)^{\frac{1}{2}} \displaystyle \frac{\Gamma\left(1-\frac{\beta}{2} \right)}{\beta \pi }\sum\limits_{\nu=0}^{2}\left(\frac{1}{ KT} \right)^{-\nu+\frac{1}{2}}\nonumber\\&\times\frac{S^{(\nu)}(0)}{\nu!}\int_{0}^{d}y^{\nu}\mathbb{E}_{\beta,\beta}\left(-y\right)e^{-xy^{-\frac{1}2}}{\rm d}y\nonumber\end{align} which when evaluated gives Meijer's $G$- function as\begin{align}
	{r}_{ij}&=n_in_j\left(\frac{8 }{\mu} \right)^{\frac{1}{2}} \displaystyle \frac{\Gamma\left(1-\frac{\beta}{2} \right)}{\beta \pi \rho }\sum\limits_{\nu=0}^{2}\left(\frac{1}{ KT} \right)^{-\nu+\frac{1}{2}}\nonumber\\&\times\frac{S^{(\nu)}(0)d^{\nu}}{\nu!z^{\nu+1}}\sum\limits_{m=0}^{\infty}\displaystyle\frac{(-1)^m(zd)^m}{\Gamma(\beta m+\beta)} \nonumber\\&\times G_{1,3}^{3,0}\begin{pmatrix}
		\frac{x^2}{4d}&\bigg\lvert
		\begin{matrix*}
		m+\nu+2 \\
			m+\nu+1,0,\frac{1}{2}
		\end{matrix*}
	\end{pmatrix}~~~~~(d<\infty).
\end{align}
 The detailed evaluation procedure of the above integral is in the Appendix.
{\section{Interpretations and concluding remarks}}
\par In this context, we undertake a comprehensive analysis of three distinct velocity distributions namely  the Maxwell-Boltzmann distribution, the pathway velocity distribution, and the  Mittag-Leffler velocity distribution.
\begin{figure}[!htb]
		\centering
		\includegraphics[width=1\linewidth]{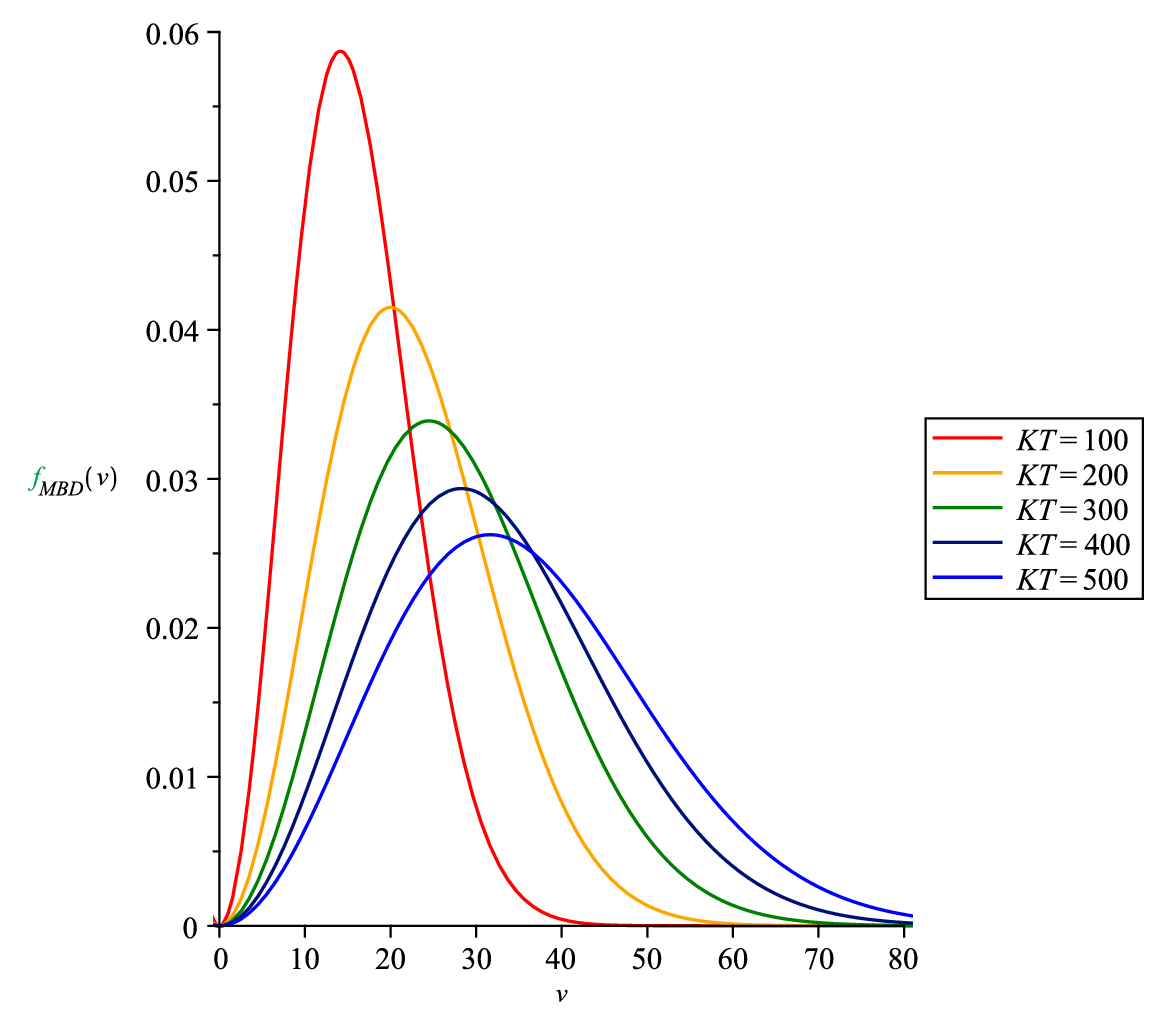}
		\caption{$f_{MBD}(v)$ for\\ $T=100,~200,~300,~400,~500.$}\label{Fig:Data1}
		\centering
		\includegraphics[width=1\linewidth]{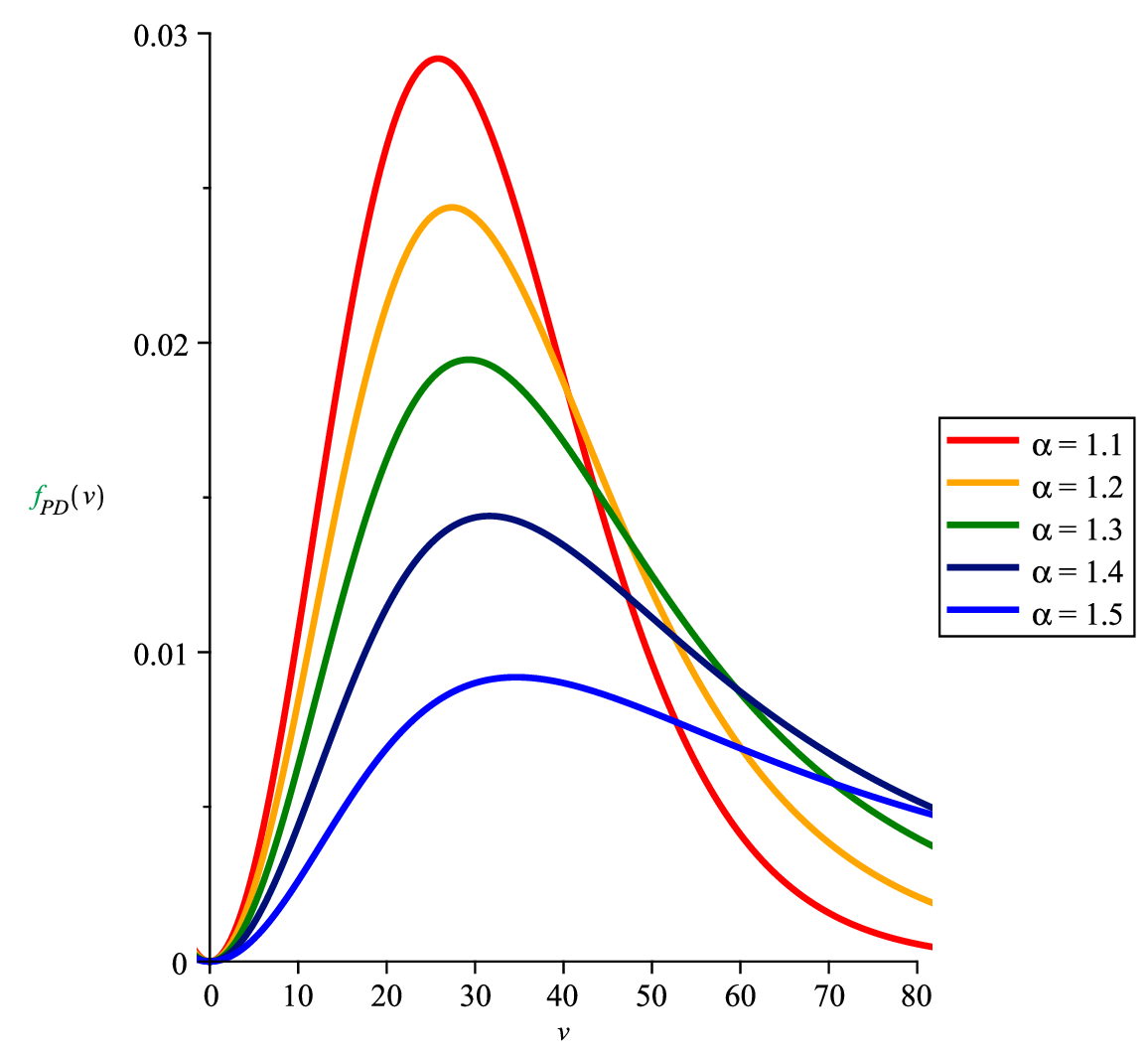}
	\caption{$f_{PD}(v)$ for $T=300$ and \\$\alpha=1.1,~1.2,~1.3,~1.4,~1.5.$}\label{Fig:Data2}
\end{figure}
\begin{figure}[!htb]
	\centering
	\includegraphics[width=1\linewidth]{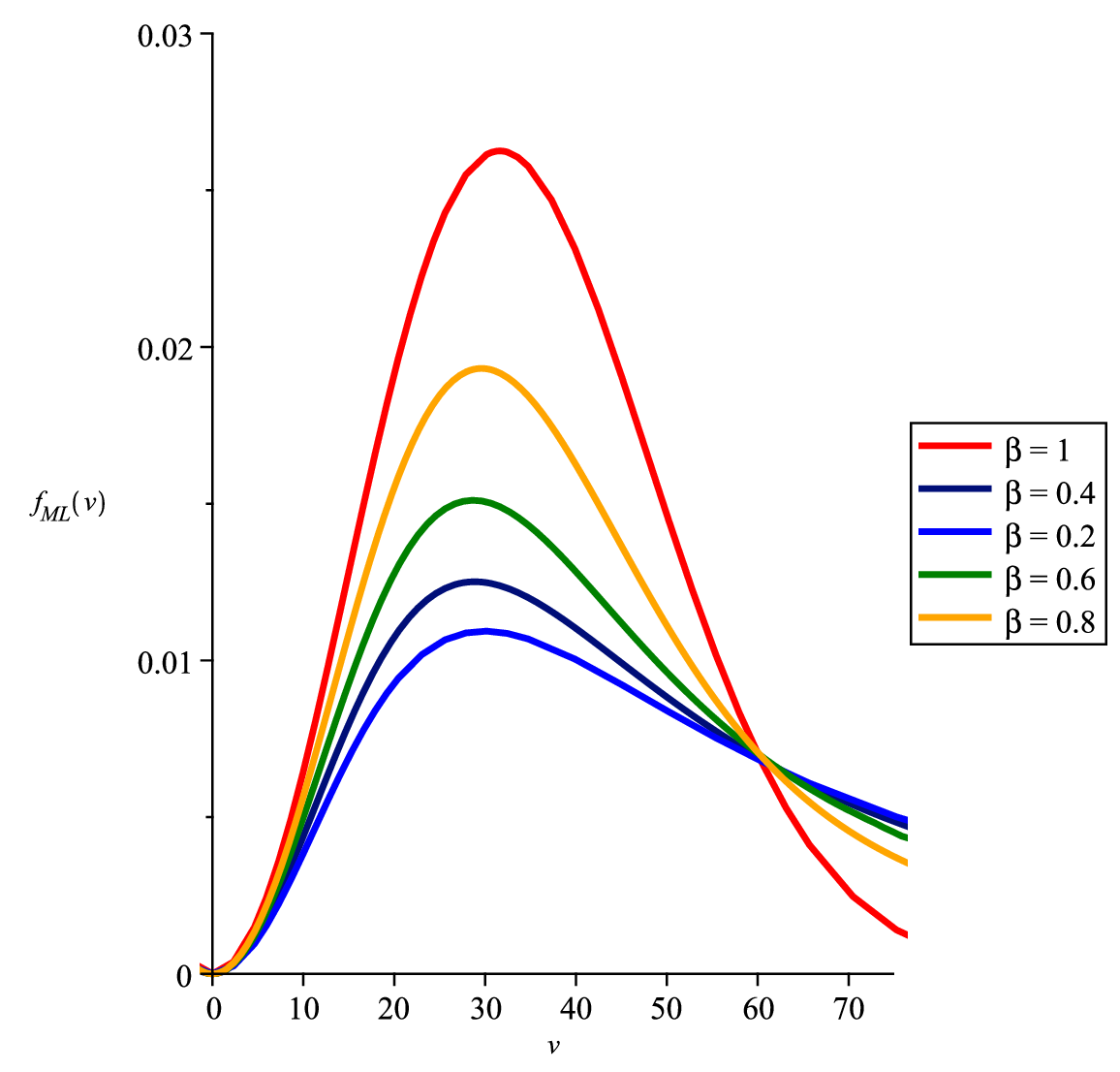}
	\caption{$f_{ML}(v)$ for $T=500$ and\\$\beta=0.2,~0.4,~0.6,~0.8,~1$}\label{Fig:Data3}
	\centering
	\includegraphics[width=1\linewidth]{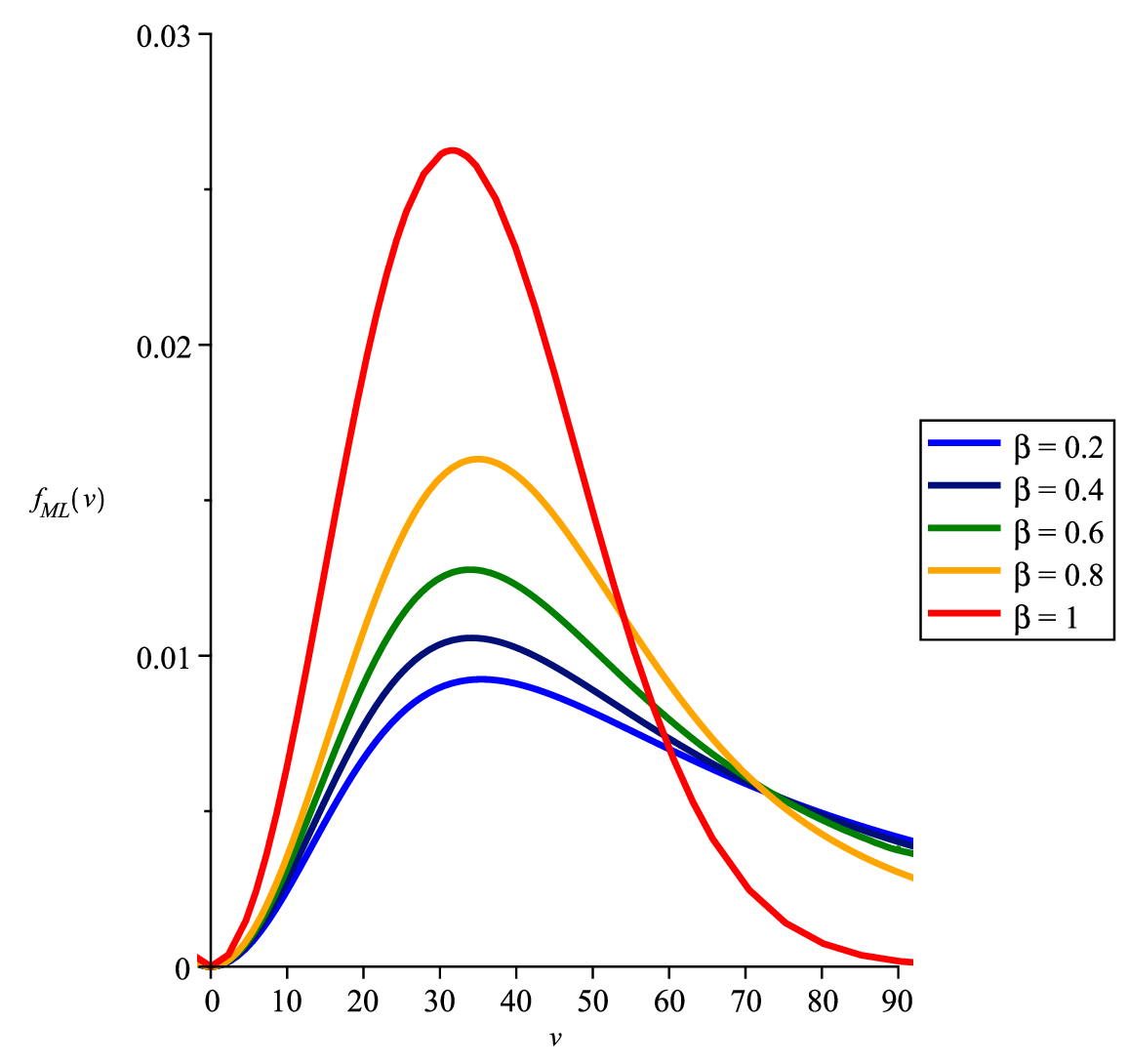}
	\caption{$f_{ML}(v)$ for $T=700$ and \\$\beta=0.2,~0.4,~0.6,~0.8,~1$}\label{Fig:Data4}
\end{figure}\par The figures below depict the velocity distribution in three different scenarios, with a reduced mass of particles, $\mu=1$. Figure \ref{Fig:Data1} illustrates $f_{MBD}(v)$ for various values of $T$, while Figure \ref{Fig:Data2} portrays the velocity distribution, $f_{PD}(v)$, under various $\alpha$ values, with $T=300$. Figure \ref{Fig:Data3} displays the Mittag-Leffler velocity distribution, for different $\beta$ values, with $T=500$, and Figure \ref{Fig:Data4} demonstrates the Mittag-Leffler velocity distribution for varying $\beta$ values, with $T=700$.  It can be observed that as the values of $T$ increases in $f_{ML}(v)$, the curve becomes heavy tailed and less peaked in comparison to $f_{PD}(v)$. Also $f_{ML}(v)$ is less peaked when compared to $f_{MBD}(v)$. It is evident from the observations that the Mittag-Leffler modified velocity distribution $f_{ML}(v)$ is more general in nature and covers many class of velocity distributions including the Maxwell-Boltzmann velocity distribution $f_{MBD}(v)$. Notably, the Maxwell-Boltzmann velocity distribution can be retrieved from the Mittag-Leffler modified velocity distribution by setting $\beta=1$. It's worth noting that the Maxwell-Boltzmann velocity distribution also serves as a limiting case of the pathway velocity distribution.
\par This study emphasises the importance of altering the velocity distribution in a fusion plasma, transitioning from the conventional Maxwell-Boltzmann distribution to pathway distribution and  modified Mittag-Leffler distribution. The study has delved into the analysis of non-resonant modified thermonuclear functions under various cases, including standard Maxwell-Boltzmann, cut-off, depleted, and screened scases. These complex functions were elegantly expressed in closed-forms, utilizing Fox's $H$-functions, showcasing the versatility and applicability of the Mittag-Leffler distribution function in plasma physics.

\section*{Appendix}
\par Consider a general integral of the form
\begin{align}
	\mathcal{T}_{1,\beta}(\nu,z,x,\rho)=\int_{0}^{\infty}y^{\nu-1}\mathbb{E}_{\beta,\beta}\left(-zy\right)e^{-xy^{-\rho}}{\rm d}y,\nonumber
\end{align}where $z>0,$ $x>0,$ $\rho>0$ and $0<\beta\leq1.$ Taking the Mellin-transform and using the definition of gamma function, we obtain,
\begin{align}\label{26}
	\mathcal{M}_{\mathcal{T}_{1,\beta}}(s)&=\displaystyle \frac{\Gamma(s)\Gamma(\nu+\rho s)\Gamma(1-\nu-\rho s)}{z^{\nu+\rho s}\Gamma(\beta-\beta\nu-\beta\rho s)},
\end{align}where $\Re(s)>0,$ $\Re(\nu+\rho s)>0.$	Using the inverse Mellin transform and equation (\ref{26}), we obtain
\begin{align}\label{21}
	&\mathcal{T}_{1,\beta}(\nu,z,x,\rho)=\displaystyle \frac{1}{z^{\nu} 2\pi i}\nonumber\\&\times\int_{\mathcal{L}}\displaystyle \frac{\Gamma(s)\Gamma(\nu+\rho s)\Gamma(1-\nu-\rho s)}{\Gamma(\beta-\beta\nu-\beta\rho s)}\left(z^{\rho} x\right)^{-s}{\rm d}s,
\end{align}where $\mathcal{L}$ is the suitable contour. Now using the Mellin-Barnes representation of $H$- function in (\ref{H}), we have\begin{align}
&\mathcal{T}_{1,\beta}({\nu},z,x,\rho)=\displaystyle \frac{1}{z^{\nu} } \nonumber\\&\times H_{1,3}^{2,1}\begin{pmatrix}
	z^{\rho}x&\bigg\lvert
	\begin{matrix*}
		(\nu,\rho) \\
		(0,1),~(\nu,\rho),~(\beta\nu,\beta \rho)
	\end{matrix*}
\end{pmatrix}.
\end{align}

\par Let us take the reaction rate integral in cut-off case to be $$\mathcal{T}_{2,\beta}^{d}(\nu,z,x,\rho)=\int_{0}^{d}y^{\nu-1}\mathbb{E}_{\beta,\beta}\left(-zy\right)e^{-xy^{-\rho}}{\rm d}y.$$ Let
\begin{align}\nonumber
	&f(x_1)=
		{x_1}^{\nu}\mathbb{E}_{\beta,\beta}\left(-zx_1\right),~~\text{if }0\leq x_1<d<\infty
\end{align}\nonumber and
 \begin{align}
	&g(x_2)=
		e^{-{x_2}^{\rho}},~~\text{if }0\leq x_2<\infty, ~\rho>0.
\end{align}\nonumber Using the Mellin convolution, we obtain
\begin{align}\label{conn}
	\mathcal{M}_{\mathcal{I}_{2,\beta}^{d}(\nu,z,x^{\rho},\rho)}(s)&=\mathcal{M}_f(s)\mathcal{M}_g(s).
\end{align}But \begin{align}
	\mathcal{M}_f(s)&=\int_{0}^{d}y^{\nu+s-1}\mathbb{E}_{\beta,\beta}\left(-zy\right){\rm d}y\nonumber\\
	&=\sum\limits_{m=0}^{\infty}\displaystyle\frac{(-1)^m(zd)^m}{\Gamma(\beta m+\beta)}\frac{d^{\nu+s}\Gamma(m+\nu+s)}{\Gamma(m+\nu+s+1)},
\end{align}and
\begin{align}
	\mathcal{M}_g(s)&=\int_{0}^{\infty}y^{s-1}e^{-y^{\rho}}{\rm d}y=\frac{1}{\rho}\Gamma\left( \frac{s}{\rho}\right), \end{align}where $\Re(s)>0$. Now, using the equation (\ref{conn}) and the inverse Mellin transform, we obtain
\begin{align}\label{i2}
	&\mathcal{I}_{2,\beta}^{d}(\nu,z,x^{\rho},\rho)=\sum\limits_{m=0}^{\infty}\displaystyle\frac{(-1)^m(zd)^m}{\Gamma(\beta m+\beta)}\frac{d^{\nu}}{2\pi i\rho}\nonumber\\&\times\int_{\mathcal{L}}\displaystyle \frac{\Gamma(m+\nu+s)\Gamma\left( \frac{s}{\rho}\right)}{\Gamma(m+\nu+s+1)}\left(\frac{x}{d} \right)^{-s} {\rm d}s,
\end{align}where $\mathcal{L}$ is the suitable contour and $\sqrt{-1}=i.$ Using (\ref{H}), obtain,
\begin{align}
	&\mathcal{I}_{2,\beta}^{d}(\nu,z,x,\rho)=\frac{d^{\nu}}{\rho}\sum\limits_{m=0}^{\infty}\displaystyle\frac{(-1)^m(zd)^m}{\Gamma(\beta m+\beta)} \nonumber\\&\times H_{1,2}^{2,0}\begin{pmatrix}
		\frac{x^{\frac{1}{\rho}}}{d}&\bigg\lvert
		\begin{matrix*}
			(m+1,1) \\
			(m,1),~(\frac{\nu}{\rho},\frac{1}{\rho})
		\end{matrix*}
	\end{pmatrix}.
\end{align}Using the duplication formula for gamma function one can see $\mathcal{I}_{2,\beta}^{d}(\nu,z,x,\rho)$ in terms of Meijer $G$-function. For detailed evaluation of the reaction rate integrals, see \cite{HJDK}

%\bibliographystyle{foo}
%\bibliography{ExampleRefs.bib}
%\bibliography{ExampleRefs}
	
\end{document}